\documentclass{sig-alternate-05-2015}
\usepackage{makecell}
\usepackage{subfigure}
\usepackage{caption}
\usepackage{color}

\usepackage{algorithm}
\usepackage[noend]{algpseudocode}

\begin{document}

\makeatletter
\def\@copyrightspace{\relax}
\makeatother

\setcopyright{acmcopyright}

%

\title{
A Supervised Learning Algorithm for Binary Domain Classification of Web Queries using {SERPs}
}

%
%
%
%
%

\numberofauthors{1} 
%
\author{
%
%
\alignauthor
Alexander C. Nwala and Michael L. Nelson\\
       \affaddr{Old Dominion University, Department of Computer Science}\\
       \affaddr{Norfolk, Virginia, 23529}\\
       \email{\{anwala, mln\}}@cs.odu.edu
}


\maketitle
\begin{abstract}
General purpose Search Engines (SEs) crawl all domains (e.g., Sports, News, Entertainment) of the Web, but sometimes the informational need of a query is restricted to a particular domain (e.g., Medical). We leverage the work of SEs as part of our effort to route domain specific queries to local Digital Libraries (DLs). SEs are often used even if they are not the ``best'' source for certain types of queries. Rather than tell users to ``use this DL for this kind of query'', we intend to automatically detect when a query could be better served by a local DL (such as a private, access-controlled DL that is not crawlable via SEs). This is not an easy task because Web queries are short, ambiguous, and there is lack of quality labeled training data (or it is expensive to create). To detect queries that should be routed to local, specialized DLs, we first send the queries to Google and then examine the features in the 
resulting Search Engine Result Pages (SERPs), and then classify the query as belonging to either the \textit{\textbf{scholar}} or \textit{\textbf{non-scholar}} domain. Using 400,000 AOL queries for the \textit{\textbf{non-scholar}} domain and 400,000 queries from the NASA Technical Report Server (NTRS) for the \textit{\textbf{scholar}} domain, our classifier achieved a precision of 0.809 and F-measure of 0.805.
\end{abstract}

%
%

%
%

%
%
\printccsdesc


\keywords{Search Engines, Digital Libraries, Web queries, Query Understanding, Web query domain classification.}

\section{Introduction}
In this paper we focus on domain classification of queries which targets two classes - the \textit{\textbf{scholar}} domain and the \textit{\textbf{non-scholar}} domain. The \textit{\textbf{scholar}} domain targets queries associated with academic or research content. For example, queries such as ``\textit{fluid dynamics}'', ``\textit{stem cell}'', ``\textit{parallel computing}'' belong to the \textit{\textbf{scholar}} domain, while queries such as ``\textit{where to find good pizza}'', ``\textit{bicycle deals}'', and ``\textit{current weather}'' belong to the \textit{\textbf{non-scholar}} domain. SEs are often used even if they
are not the ``best'' source for certain types of queries. For example, according to the Ithaka S+R Faculty Survey 2015 \cite{survey}, faculty members are likely to begin their research discovery with a general purpose SE. In fact, according to this survey, over 30\% of the surveyed faculty report SEs as the start point of their academic research discovery \cite{survey}. Rather
than tell users to ``use this DL for this kind of query'', we
intend to automatically detect when a query could be better
served by a local DL (such as a private, access-controlled
DL that is not crawlable via SEs). We propose a novel method which does not rely on processing the actual query. Instead, we trained a classifier based on the features found in a Google SERP (Search Engine Result Page). The classifier was trained and evaluated (through 10-fold cross validation) on a dataset of 600,000 SERPs evenly split across both classes. The results were validated by 200,000 SERPs evenly split across both classes yielding a classification precision of 0.809 and F-measure of 0.805. We targeted a binary class, however our method could be scaled to accommodate other classes if the right features are found.

The rest of the paper is organized as follows: in Section 2, we present research similar to ours but outline how our work differs from the state of the art. Section 3 gives an overview of our solution in two stages: building the classifier and classifying a query. In Section 4, we identify the features and discuss the rationale for selecting the features. Section 5 outlines the classifier building/evaluation and classification tasks. We present our results in Section 6, discuss our findings in Section 7, and finally in
Section 8 we conclude.

\section{RELATED WORK}

The problems of query domain classification and query routing to specialized collections have been studied extensively. Jingbo et al. \cite{Jingbo:label} built a domain knowledge base from web pages for query classification. Gravano, et al. \cite{Gravano:reasoning} built a classifier targeting the geographical locality domain. Dou Shen et al. \cite{shen2005q} built an ensemble-search based approach for query classification in which queries are enriched with information derived from Search Engines. Wang et al. \cite{wang2009semi} in an effort to label queries, investigated the use of 
semi-supervised learning algorithms built from HTML lists to automatically generate semantic-class lexicons. For the query understanding task (which includes domain classification), Liu et al. \cite{liu2013query} proposed a method which takes advantage of the linguistic information encoded in hierarchical parse trees for query understanding. This was done by extracting a set of syntactic structural features and semantic dependency features from query parse trees. Jansen et al. \cite{jansen2008determining} presented a classification of user intent. Three hierarchical intent classes were considered: informational, navigational, and transactional intent.

Sugiura and Etzioni \cite{sugiura2000query} introduced Q-pilot: a system which dynamically routes user queries to their respective specialized search engines. Similarly, in order to improve search results, Lee et al. \cite{lee2005automatic}
proposed two types of features (user-click behavior and anchor-link distribution) in order to discover the ``goal'' behind a user's Web query. Jian Hu et al. \cite{hu2009understanding} proposed a method of discovering a user query's intent with Wikipedia in order to help a Search Engine to automatically route the query to some appropriate Vertical Search Engines. In this work, the Wikipedia concepts were used as the intent representation space (each intent domain is represented as a set of Wikipedia articles and categories). Thereafter, the intent of incoming queries were identified through mapping the query into the Wikipedia representation space.

The query domain classification problem is not new, and discovering the domain of a query facilitates routing the query to an appropriate Search Engine or Digital Library. However, since queries are short, ambiguous, and in constant flux, maintaining a labeled training dataset is expensive. Similar to other methods, our method utilized the Google Search Engine, but differs from the state of the art by not processing the query directly, but the SERP instead. We used Google SERPs to develop our method. However, our techniques could be applied to other Search Engines such as Bing, once the corresponding features extracted from Google SERPs are identified on the Bing SERPs. Fig. 8 identifies some features on the Bing SERP which we collect from Google SERPs.

\begin{figure*}
	\centering
	\includegraphics[height=0.8\textheight, width=\textwidth]{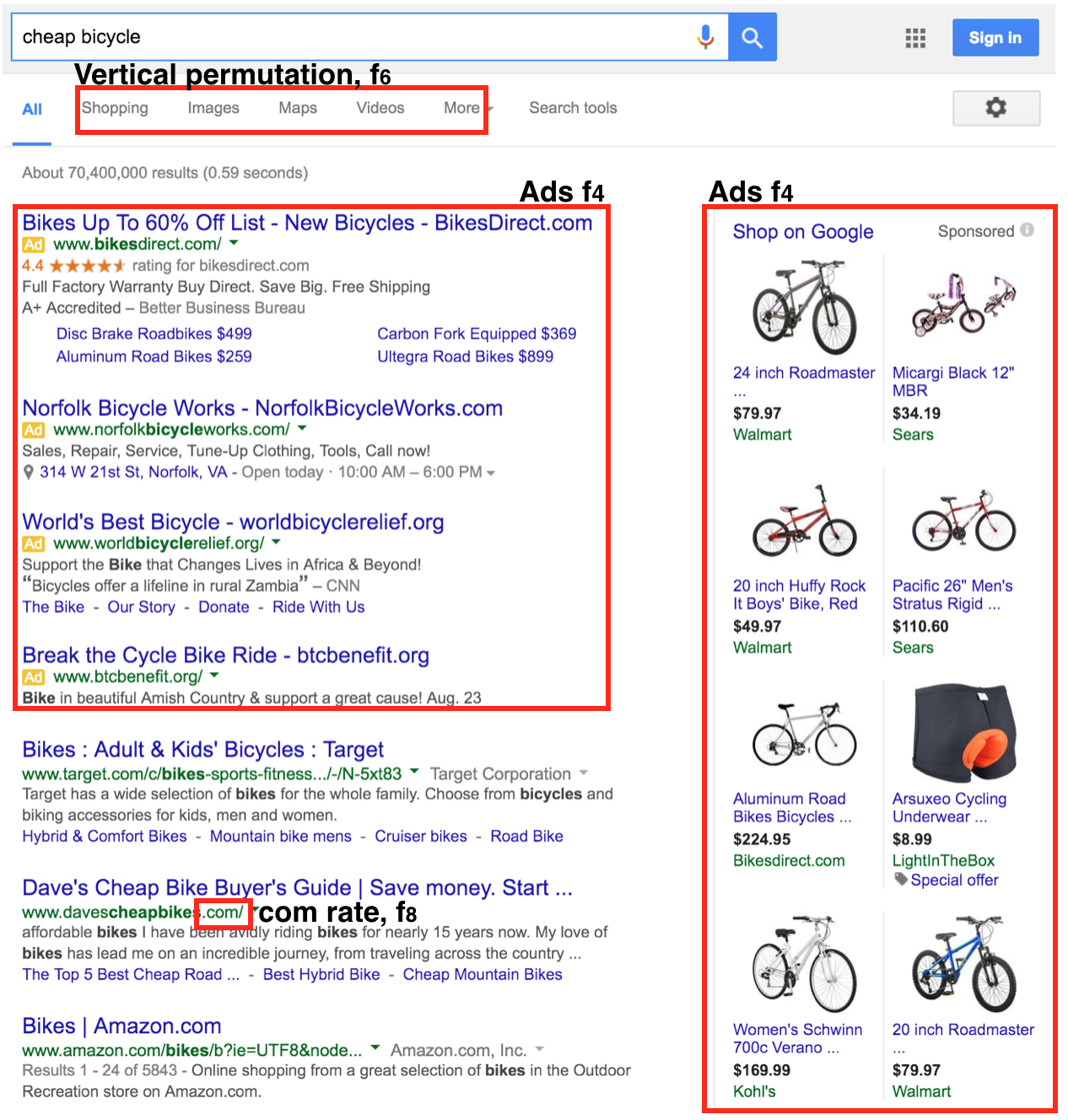}
	\caption{\textit{\textbf{non-scholar}} queries typically feature ``shopping'' in the Vertical permutation ($f_6$), Ads 
($f_4$), and are dominated by \textit{com} TLDs ($f_8$). Snapshot date: March 15, 2016.}
\end{figure*}

\begin{figure*}
	\centering
	\includegraphics[height=0.8\textheight, width=0.7\textwidth]{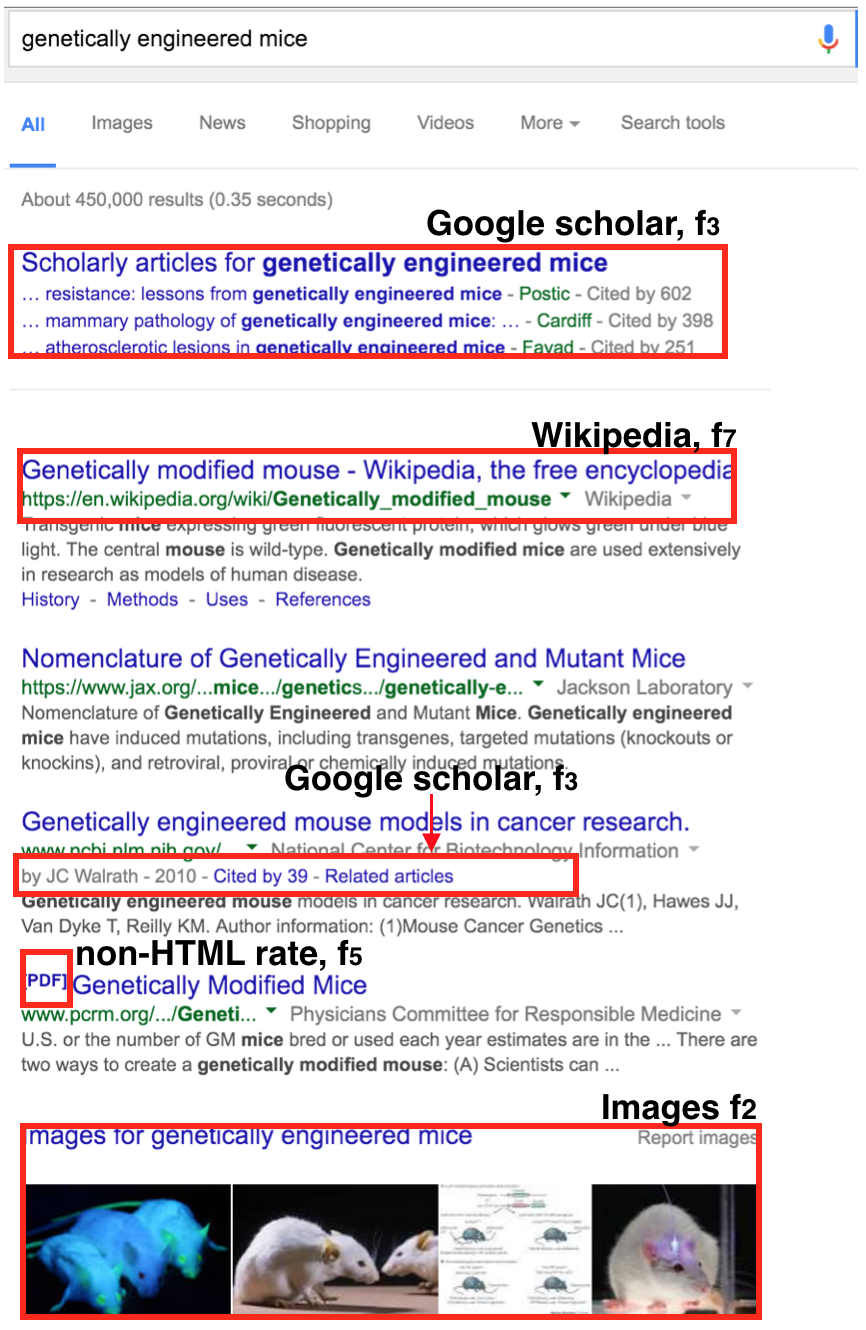}
	\caption{\textit{\textbf{scholar}} queries typically feature results from Wikipedia ($f_7$) and Google 
Scholar ($f_3$), as well as \textit{non-html} types ($f_5$). Note: Google feedback panel was removed from image to accommodate more features. Snapshot date: March 15, 2016.}
\end{figure*}

\begin{table*}
\centering
\caption{Complete list of features}
\label{tab:completeListFeatures}
\begin{tabular}{|l|c|c|l|} \hline 
\textbf{Feature} & \textbf{Domain} & \textbf{Information Gain} & \textbf{Gain Ratio} \\ \hline
$f_1$, Knowledge Entity & \{0, 1\} & 0.013 & 0.033 \\ \hline
$f_2$, Images & \{0, 1\}  & 0.021 & 0.024 \\ \hline
$f_3$, Google Scholar & \{0, 1\}  & 0.248 & 0.298 \\ \hline
$f_4$, Ad ratio & ${\rm I\!R}\in[0, 1]$  & 0.051 & 0.022 \\ \hline
$f_5$, \textit{non-HTML} rate & ${\rm I\!R}\in[0, 1]$  & 0.205 & 0.048 \\ \hline
$f_6$, Vertical permutation & $\mathbb{Z}\in[1, 336]$  & 0.100 & 0.020 \\ \hline
$f_7$, Wikipedia & \{0, 1\}  & 0.033 & 0.034 \\ \hline
$f_8$, \textit{com} rate & ${\rm I\!R}\in[0, 1]$  & 0.127 & 0.092 \\ \hline
$f_9$, Max. Title Dissimilarity & ${\rm I\!R\in[0, 1]}$  & 0.018 & 0.007 \\ \hline
$f_{10}$, Max. Title Overlap & $\mathbb{Z}\in[0, \max_{t_i}| q' \cap t_i |]$  & 0.031 & 0.013 \\ \hline
\end{tabular}
\end{table*}

\begin{table*}
\centering
\caption{Proportion of SERPs with a Knowledge Entity ($f_1$), Images ($f_2$), Google Scholar Citation ($f_3$), and a Wikipedia link ($f_7$), for \textit{\textbf{scholar}} and \textit{\textbf{non-scholar}} classes for 200,000 SERPs evenly split across both classes.}
\label{tab:percentPresentAbsent}
\begin{tabular}{|l|c|l|} \hline 
\textbf{Feature Title} & \textbf{Scholar} & \textbf{non-Scholar} \\ \hline
Knowledge Entity ($f_1$) & 4.10\% & 10.97\% \\ \hline
Images ($f_2$) & 38.27\% & 22.60\% \\ \hline
Google Scholar ($f_3$) & 50.51\% & 2.54\% \\ \hline
Wikipedia ($f_7$) & 49.63\% & 28.72\% \\ \hline
\end{tabular}
\end{table*}

\begin{table*}
\centering
\caption{Summary Statistics for Ad ratio ($f_4$), \textit{non-HTML} rate ($f_5$), Vertical permutation ($f_6$), \textit{com} rate ($f_8$), Max. Title Dissimilarity ($f_9$) and Max. Title Overlap ($f_{10}$), for \textit{\textbf{scholar}} and \textit{\textbf{non-scholar}} classes for 200,000 SERPs evenly split across both classes.}
\label{tab:summaryStats}
\begin{tabular}{|l|c|c|c|c|c|}
\multicolumn{1}{r}{} & \multicolumn{2}{c}{\textbf{Scholar}} & \multicolumn{2}{c}{\textbf{non-Scholar}} \\ \hline
\textbf{Feature Title} & \textbf{Mean} & \textbf{Std. Dev} & \textbf{Mean} & \textbf{Std. Dev} \\ \hline
Ad ratio ($f_4$) & 0.0470 & 0.1100 & 0.1060 & 0.1680 \\ \hline
\textit{non-HTML} rate ($f_5$) & 0.1660 & 0.1910 & 0.0170 & 0.0630 \\ \hline
Vertical permutation ($f_6$) & 228.9220 & 59.8090 & 240.3640 & 56.4760 \\ \hline
\textit{com} rate ($f_8$) & 0.2740 & 0.2230 & 0.1000 & 0.1790 \\ \hline
Max. Title Dissimilarity ($f_9$) & 0.8420 & 0.0870 & 0.8250 & 0.0870 \\ \hline
Max. Title Overlap ($f_{10}$) & 1.8570 & 1.1660 & 1.8810 & 1.4630 \\ \hline
\end{tabular}
\end{table*}

\section{Method Overview of Learning Algorithm for Domain Classification}

Our solution to the  web query binary domain classification problem can be summarized in two stages:

\begin{enumerate}
\item \textbf{Stage 1 - Building the classifier:} 

First, identify the discriminative features. Second, build a dataset for the \textit{\textbf{scholar}} domain class and \textit{\textbf{non-scholar}} domain class. Third, train a classifier. Fourth, evaluate the classifier using 10-fold cross validation.

\item \textbf{Stage 2 - Classifying a query:} 

First, issue the query to Google and download the SERP. Second, extract the features. Finally, use the classifier hypothesis function to make a prediction.
\end{enumerate}

\section{Extracting Features from SERPs:}After extensive study (information theory and visualization), we identified 10 features to be extracted from the Google SERP:

\begin{enumerate}
	\item \textbf{Knowledge Entity ($f_1$):}

	\textit{Domain:}
	Binary, $f_1 \in \{0, 1\}$

	\textit{Figure Example when Present:}
	Fig. 3

	\textit{Figure Example when Absent:}
	Fig. 1

	\textit{Description:}
	$f_1$ indicates the presence or absence of the Google Knowledge Graph Panel. The presence of this panel (Fig. 3) often indicates the presence of a real world object such as a person, place, etc. as understood by Google.

	\begin{figure*}
	\centering
	\includegraphics[height=0.75\textheight, width=\textwidth]{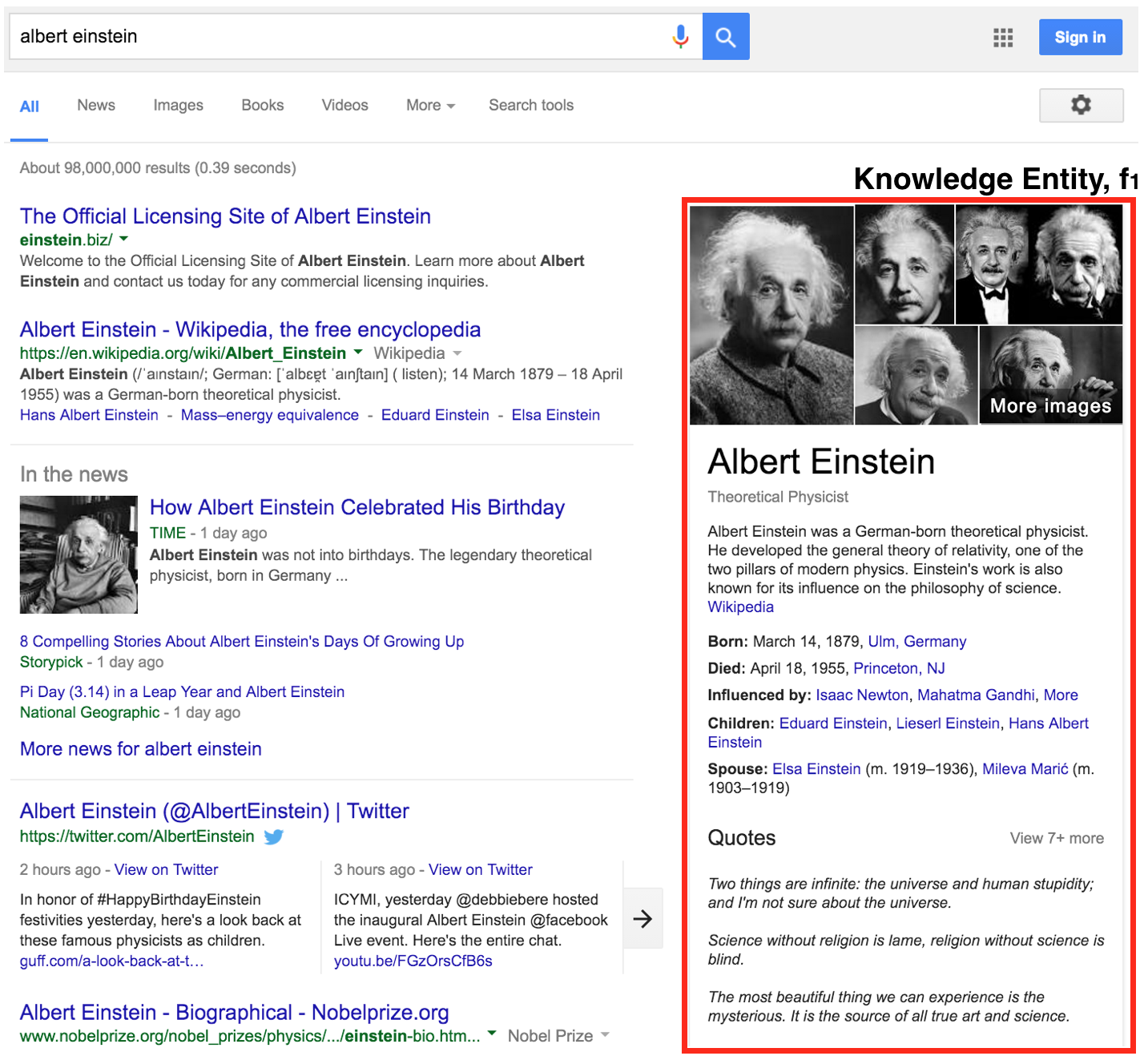}
	\caption{Google Knowledge Graph Panel ($f_1$) for ``albert einstein'' query. Snapshot date: March 15, 2016.}
	\end{figure*}

	\item \textbf{Images ($f_2$):}

	\textit{Domain:}
	Binary, $f_2 \in \{0, 1\}$

	\textit{Figure Example when Present:}
	Fig. 2

	\textit{Figure Example when Absent:}
	Fig. 1

	\textit{Description:}
	$f_2$ indicates the presence or absence of Images on the SERP.

	\item \textbf{Google Scholar ($f_3$):}

	\textit{Domain:}
	Binary, $f_3 \in \{0, 1\}$

	\textit{Figure Example when Present:}
	Fig. 2

	\textit{Figure Example when Absent:}
	Fig. 1

	\textit{Description:}
	$f_3$ indicates the presence or absence of a citation. A citation of a research paper or book is often a strong indicator about a web query of the \textit{\textbf{scholar}} class (Table ~\ref{tab:percentPresentAbsent}).

	\item \textbf{Ad ratio ($f_4$): }

	\textit{Domain:}
	${\rm I\!R}$, $f_4 \in [0, 1]$

	\textit{Figure Example when Present:}
	Fig. 1

	\textit{Figure Example when Absent:}
	Fig. 2

	\textit{Description:}
	$f_4$ represents the proportion of ads on the SERP. Even though Search Engines strive to include ads (in order to pay their bills), our information analysis showed that the \textbf{\textit{non-scholar}} class SERPs had the higher proportion of ads (Fig. 4).

	\begin{figure*}
	\centering
	\includegraphics[height=4in, width=4in]{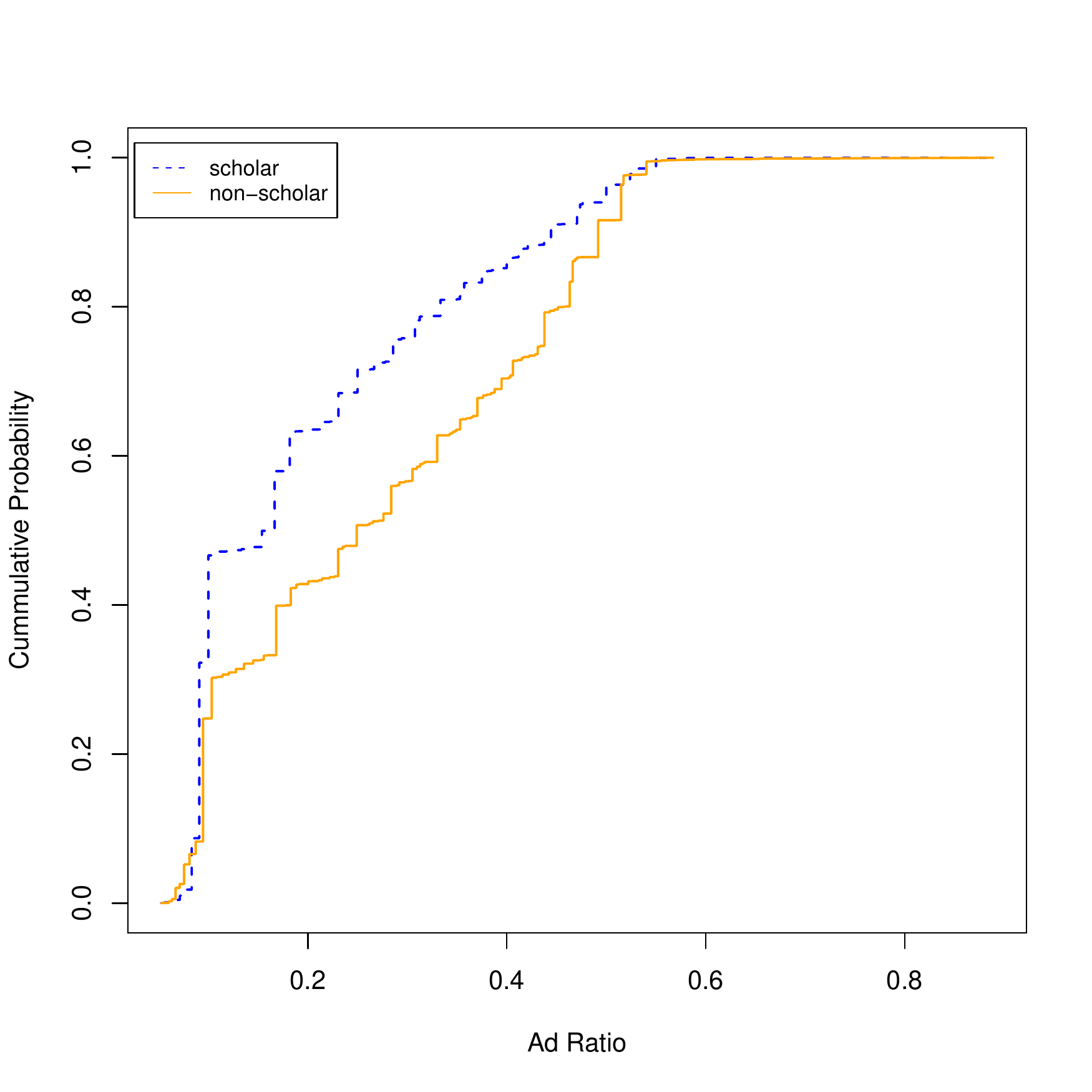}
	\caption{CDF of Ad Ratio ($f_4$) for \textbf{\textit{scholar}} (blue, dashed) \& \textbf{\textit{non-scholar}} (orange, solid) classes for 200,000 SERPs evenly split across both classes; \textbf{\textit{non-scholar}} with a higher Probability of ads.}
	\end{figure*}

	\begin{figure*}
	\centering
	\includegraphics[height=4in, width=4in]{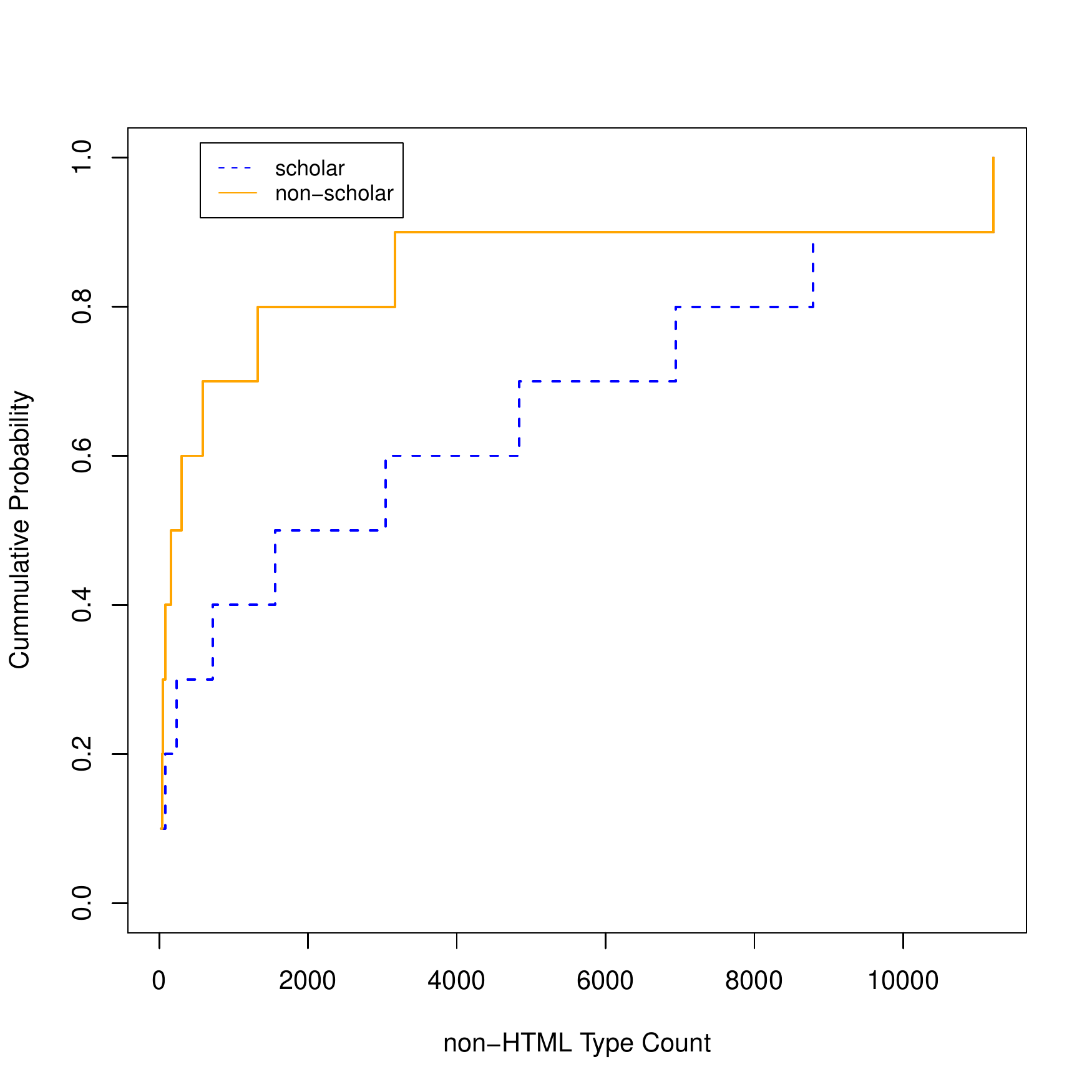}
	\caption{CDF of \textit{non-html} types ($f_5$), \textit{\textbf{scholar}} (blue, dashed) and \textit{\textbf{non-scholar}} (orange, solid) classes for 200,000 SERPs evenly split across both classes; \textit{\textbf{scholar}} with a higher probability of \textit{non-html} types.}
	\end{figure*}

	\item \textbf{\textit{non-HTML} rate ($f_5$):}

	\textit{Domain:}
	${\rm I\!R}$, $f_5 \in [0, 1]$

	\textit{Figure Example when Present:}
	Fig. 1 and Fig. 2

	\textit{Figure Example when Absent:}
	NA

	\textit{Description:}
	$f_5$ represents proportion of \textit{non-html} types on the SERP as reported by Google (as opposed to dereferencing the links and retrieving the ``Content-Type'' header values). We defined \textit{non-html} types to include members of set \textbf{M}.
	
	\textbf{M} \textit{= \{pdf, ppt, pptx, doc, docx, txt, dot, dox, dotx, rtf, pps, dotm, pdfx\}}

	For example, if a given SERP has results with the following Mime types:

	\begin{verbatim}
	1.  html 
	2.  html
	3.  html
	4.  pdf  (non-html)
	5.  html
	6.  ppt  (non-html)
	7.  html
	8.  pptx (non-html)
	9.  html
	10. html
	\end{verbatim}

	Since there are three \textit{non-html} types, $f_5 = \frac{3}{10}$

	Our intuition which expects the \textit{\textbf{scholar}} class to have more \textit{non-html} types was matched by Fig. 5.

	\item \textbf{Vertical permutation ($f_6$):}

	\textit{Domain:}
	$\mathbb{Z}$, $f_6 \in [1, 336]$

	\textit{Figure Example when Present:}
	Fig. 1 and Fig. 2

	\textit{Figure Example when Absent:}
	NA

	\textit{Description:}
	$f_6$ represents the rank from $^{8}P_{3}$ (8 Permutation 3) possible page order of the SERP.
	In addition to the default SERP (All), the Google SERP provides a set of eight custom vertical search pages: 
	\begin{verbatim}
	Apps, Books, Flights, Images, Maps, News,
	Shopping, and Videos
	\end{verbatim}
	Even though there are eight possible vertical pages, only four are displayed simultaneously with the default SERP (All). The rest are hidden. Different queries trigger different permutations of four from the set of eight. The order in which the four pages are displayed is tied to the kind of query. 

	For example, the query ``\textit{cheap bicycle}'' (Fig. 1) presents:
	\begin{verbatim}
	Shopping, Images, Maps, and Videos
	\end{verbatim}
	And the query ``\textit{genetically engineered mice}'' (Fig. 2) presents:
	\begin{verbatim}
	Images, News, Shopping, and Videos
	\end{verbatim}
	In order to capture this information we assigned each page a number based on its position in alphabetical order:
	\begin{verbatim}
	0 - Apps, 1 - Books, 2 - Flights, 3 - Images,
	4 - Maps, 5 - News, 6 - Shopping, 7 - Videos
	\end{verbatim}
	We empirically determined that the relevance of a query to the vertical pages decreases as we go from the first vertical page to the last vertical page. For example, the query ``\textit{cheap bicycle}'' is most relevant to the \textit{Shopping} vertical page, and less relevant to the \textit{Images} vertical page and so on. Due to the decrease in relevance we decided to capture only the first three pages. This gives a total of 336 ($^{8}P_{3}$) permutations (in lexicographic order) beginning from Rank 0 to Rank 335:
	
	\begin{verbatim}
	Permutation 0:   0 1 2: Apps, Books, Flights
	Permutation 1:   0 1 3: Apps, Books, Images
	Permutation 2:   0 1 4: Apps, Books, Maps
	.
	.
	.
	Permutation 335: 7 6 5: Videos, Shopping, News
	\end{verbatim}

	\item \textbf{Wikipedia ($f_7$):}

	\textit{Domain:}
	Binary, $f_7 \in \{0, 1\}$

	\textit{Figure Example when Present:}
	Fig. 2

	\textit{Figure Example when Absent:}
	Fig. 1

	\textit{Description:}
	$f_7$ indicates the presence or absence of a Wikipedia link on the SERP. Based on our study, the \textit{\textbf{scholar}} class presented a higher proportion of SERPs with a Wikipedia link (Table ~\ref{tab:percentPresentAbsent}). 

	\item \textbf{\textit{com} rate ($f_8$):}

	\textit{Domain:}
	${\rm I\!R}$, $f_8 \in [0, 1]$

	\textit{Figure Example when Present:}
	Fig. 1 and Fig. 2

	\textit{Figure Example when Absent:}
	NA

	\textit{Description:}
	$f_8$ represents the proportion of \textit{com} links on the SERP. The motivation for this feature was due to observations from visualizing our datasets which revealed that queries of the \textit{\textbf{scholar}} class had a lesser proportion of \textit{com}  links on the SERP, instead had a higher proportion of TLDs such as \textit{edu} and \textit{gov}.

	\item \textbf{Maximum Title Dissimilarity ($f_9$):}

	\textit{Domain:}
	${\rm I\!R}$, $f_9 \in [0, 1]$

	\textit{Figure Example when Present:}
	Fig. 1 and Fig. 2

	\textit{Figure Example when Absent:}
	NA

	\textit{Description:}
	$f_9$ represents the maximum dissimilarity value between the query and all SERP titles:
	Given a query $q$, with SERP title $t_i$, longest SERP title $T$, Levinshtein Distance function $LD$, and cardinality (length) of longest SERP title $|T|$, then:

	\begin{equation}
		f_9 = \max_{t_i}\frac{LD(q, t_i)}{|T|}
	\end{equation}

	For example, considering the query ``cheap bicycle'' in Fig. 1, the maximum $f_9$ (dissimilarity) score is 0.8182, which belongs to the second title - ``Dave's Cheap Bike Buyer's Guide | Save money. Start ...''. For the query ``genetically engineered mice'' (Fig. 2), the maximum $f_9$ is 0.6557 which belongs to the first title - ``Genetically modified mouse - Wikipedia, the free encyclopedia''.
	See Section 5.3 for an example of how $f_9$ was calculated for a given query.

	\item \textbf{Maximum Title Overlap ($f_{10}$):}

	\textit{Domain:}
	$\mathbb{Z}$, $f_{10} \in \mathbb{Z}\in[0, \max_{t'_i}| q' \cap t'_i |]$

	\textit{Figure Example when Present:}
	Fig. 1 and Fig. 2

	\textit{Figure Example when Absent:}
	NA

	\textit{Description:}
	In contrast with $f_9$ which captures dissimilirity, $f_{10}$ represents the cardinality of the maximum common value between the query and all SERP titles:
	Given a query set $q'$ with SERP title set $t'_i$, then:

	\begin{equation}
		f_{10} = \max_{t'_i} | q' \cap t'_i |
	\end{equation}
	For example, for the query ``cheap bicycle'' (Fig. 1), the maximum $f_{10}$ score is 1 which belongs to the second title - ``Dave's Cheap Bike Buyer's Guide | Save money. Start ...''. For the query ``genetically engineered mice'' (Fig. 2), the maximum $f_{10}$ score is 3 which belongs to the second title - ``Nomenclature of Genetically Engineered and Mutant Mice''. 
	See Section 5.3 for an example of how $f_{10}$ was calculated for a given query.

\end{enumerate}

Individually, many features may not be helpful in distinguishing the \textit{\textbf{scholar}} query from the \textit{\textbf{non-sholar}} query, however it is in the combination of these features in which we could learn to classify web queries based on the combined signals these features indicate.

Figures 1 and 2 present a subset of the features for both classes while Table ~\ref{tab:completeListFeatures} presents the complete list of features. Note the ads ($f_4$) present in the \textit{\textbf{non-scholar}} SERP and the absence of PDF ($f_5$) documents (Fig. 1). For the \textit{\textbf{scholar}} SERP (Fig. 2), note the presence of a Wikipedia link ($f_7$), the PDF ($f_5$) document and the Google Scholar article ($f_3$). Therefore, at scale, we could learn from not just the presence of a feature, but its absence.

\section{Classifier Training and Evaluation:}
After feature identification, we downloaded the Google SERPs for 400,000 AOL 2006 queries \cite{Pass:2006:PS:1146847.1146848} and 400,000 NTRS (from 1995-1998) queries \cite{ntrs:ir} for the non-Scholar and Scholar datasets, respectively. We thereafter extracted the features described in Section 4, and then trained a classifier on the dataset. After exploring multiple statistical machine learning methods such as SVMs, we chose logistic regression because of its fast convergence rate and simplicity.
\subsection{Logistic Regression:}
Logistic regression is analogous to linear regression.
Linear regression describes the relationship between a continuous response variable and a set of predictor variables, but logistic regression approximates the relationship between a discrete response variable and a set of predictor variables. The description of logistic regression in this section is based on Larose's explanation \cite{larose2006data}.
\subsubsection{Logistic Regression Hypothesis Representation:}
Given a predictor variable $X$, and a response variable $Y$, the logistic regression hypothesis function (outputs estimates of the class probability), is the conditional
mean of $Y$ given $X = x$, is denoted as $P(Y|x)$. The hypothesis function represents the expected value of the response variable for a given value of the predictor and is outlined by Eqn. 3

\begin{equation}
P(Y|x) = 
\frac{e^{\beta_0+\beta_1x}}{1 + e^{\beta_0+\beta_1x}}
;0 \leq P(Y|x) \leq 1
\end{equation}

$P(Y|x)$ is nonlinear and sigmoidal (S-shaped), which is unlike the linear regression model which assumes the relationship between the predictor and response variable is linear. Since $0 \leq P(Y|x) \leq 1$, $P(Y|x)$ may be interpreted as a probability. For example, given a two class problem where $Y$ corresponds to the presence of a disease ($Y=1$) or the absence of the same disease ($Y=0$),  

$P(Y=1|x) + P(Y=0|x) = 1$. Thus $P(Y|x)$ may be threshold at $0.5$: 

If $P(Y|x) \geq 0.5;$ Y = 1

If $P(Y|x) < 0.5;$ Y = 0

\subsubsection{Logistic Regression Coefficients Estimation - Maximum Likelihood Estimation:}
The least-squares method offers a means of finding a closed-form solution for the optimal values of the regression coefficients for linear regression. But this is not the case for logistic regression - there is no such closed-form solution for estimating logistic regression coefficients. Therefore, we may apply the \textit{maximum likelihood
estimation} method to find estimates of the parameters $\beta_i$ which maximize the likelihood of observing the data. Consider the likelihood function $l(\beta|x)$ (Eqn. 4), 

Let $P(Y|x) = \alpha(x)$
\begin{equation}
l(\beta|x) = \prod_{i=1}^{n}[\alpha(x_i)]^{y_i}[1 - \alpha(x_i)]^{1 - y_i}
\end{equation}

which represents the parameters $\beta_i$ that describe the probability of the observed data, $x$. To discover the maximum likelihood estimators, we must find the values of the parameters $\beta_i$ ($\beta_0, \beta_1, \beta_2,...,\beta_m)$, which maximize the likelihood function $l(\beta|x)$.

The log likelihood of $l(\beta|x)$, $L(\beta|x)$ is often preferred to the likelihood function, because it is more computationally tractable, thus expressed by Eqn. 5:

\begin{equation}
L(\beta|x) = ln [l(\beta|x)] = \sum_{i=1}^{n} \{y_iln[\alpha(x_i)] + (1-y_i)ln[1-\alpha(x_i)]\}
\end{equation}

To find the \textit{maximum likelihood estimators} we have to differentiate $L(\beta|x)$ with
respect to each parameter. Thereafter, we set the result equal to zero and solve the equation. For linear regression a closed-form solution for the differentiation exists, but this is not the case for logistic regression, so me must resort to other numerical methods such as \textit{iterative weighted
least squares} 
\cite{mccullagh1989generalized}

\subsection{Weka Logistic Regression:}
Using Weka \cite{hall2009weka}, we built a Logistic Regression Model (Eqns. 6, 7) on a 600,000 dataset evenly split across both classes. The model was evaluated using 10-fold cross validation yielding a classification precision of   0.809 and F-Measure of 0.805

\begin{equation}
p_q = 
\frac{e^{g_q}}{1 + e^{g_q}}
\end{equation}
\begin{equation}
g_q = 2.7585 + \sum_{i=1}^{10}c_i f_i
\end{equation}

$c_i\in C$ : The coefficient of feature $f_i$.

$C$: The coefficients matrix output by Weka contains the coefficients $c_i$ for each feature $f_i$.

\[
C=
\begin{bmatrix}
    c_1 \\
    c_2 \\
    c_3 \\
    c_4 \\
    c_5 \\
    c_6 \\
    c_7 \\
    c_8 \\
    c_9 \\
    c_{10}
\end{bmatrix}
=
\begin{bmatrix}
    0.8266 \\
    -1.1664 \\
    -2.7413 \\
    -1.7444 \\
    6.2504 \\
    -0.0017 \\
    -1.0145 \\
    -1.5367 \\
    1.8977 \\
    -0.1737
\end{bmatrix}
\]

To find the domain $d\in$ \{Scholar, non-Scholar\} of a query $q$, download Google SERP, then apply Algorithm 1.

\begin{algorithm}
\caption{Classify Query}
\label{alg:euclid}

\textbf{Input:} Google SERP $s$ and query, $q$

\textbf{Output:} \textit{scholar} or \textit{non-scholar} domain label for $q$
\begin{algorithmic}
\Function{ClassifyQuery}{$s$, $q$}\Comment{SERP $s$ for query $q$} 
\State 1. $f_1$ ... $f_{10} \gets s$\Comment{Initialize features}
\State 2. $g_q = 2.7585 + \sum_{i=1}^{10}c_i f_i$\Comment{Eqn 3; estimate $g_q$}

\State 3. \large $p_q = 
\frac{e^{g_q}}{1 + e^{g_q}}$\normalsize \Comment{Eqn 4; estimate class probability $p_q$}
\newline
\If{$p_q \geq 0.5$} \Comment{Set domain}
\State 4. \textbf{return} \textit{scholar domain}
\Else
\State 5. \textbf{return} \textit{non-scholar domain}
\EndIf
\EndFunction
\end{algorithmic}
\end{algorithm}

\subsection{Classifying a Query (Example):}

Consider the process of classifying the query: ``\textit{moon shot}'' based on Algorithm 1. This example references Fig. 7. Also for this example, when a binary feature is present, we represent the feature value with 0 and when it is absent, we represent the feature value with 1. This is done in order to conform to Weka's convention, such that the result is comparable with Weka's result.

\begin{table}
\centering
\caption{$f_9$ and $f_{10}$ calculated for SERP Titles for Query ``\textit{moon shot}''. Note that titles $t_1$ and $t_2$ are similar but not exact.}
\label{table:f9f10}
\begin{tabular}{c|c|c|c} \hline 
& & \\
$t_i$ & \textbf{Title} & $f_9$ & $f_{10}$ \\ & & \\ \hline \hline
$t_1$ & \makecell{Moon Shot - Wikipedia,\\ the free encyclopedia} & 0.7955 & 2 \\ \hline
$t_2$ & \makecell{Moonshot - Wikipedia,\\ the free encyclopedia} & 0.8372 & 0 \\ \hline
$t_3$ & \makecell{HP Moonshot System\\ | HP Official Site} & 0.7568 & 0 \\ \hline
$t_4$ & \makecell{What is moonshot?\\ - Definition from WhatIs.com} & 0.8478 & 0 \\ \hline
$t_5$ & \makecell{Thinking big: Larry Page\\ talks moon shots - Google Careers} & 0.8448 & 1  \\ \hline
$t_6$ & Moon Shots Program & 0.5000 & 1  \\ \hline
$t_7$ & Moon Shot - Amazon.com & 0.5909 & 2  \\ \hline
$t_8$ & \makecell{Moonshot!, John \\Sculley - Amazon.com} & 0.8056 & 0 \\ \hline
$t_9$ & \makecell{Google's Larry Page on\\ Why Moon Shots Matter} & 0.7955 & 1 \\ \hline
$t_{10}$ & \makecell{Google's Moon Shot\\ - The New Yorker} & 0.7429 & 2 \\ \hline
$t_{11}$ & Moon Shots for Management & 0.6400 & 1 \\ \hline \hline
& \textbf{Maximum} & 0.8478 & 2 \\ \hline
\end{tabular}
\end{table}

\begin{table*}
\centering
\caption{Detailed Accuracy by Class for Logistic Regression Model}
\label{table:detailedAccurayLogistic}
\begin{tabular}{|c|c|c|c|c|c|c|c|} \hline 
\textbf{Class} & \textbf{TP Rate} & \textbf{FP Rate} & \textbf{Precision} & \textbf{Recall} & \textbf{F-measure} & \textbf{ROC Area} &  \\ \hline
\textbf\textit{scholar} & 0.748  &   0.138  &    0.845  &   0.748  &   0.793   &   0.890 & \\ \hline
\textbf\textit{non-scholar} & 0.862  &   0.252  &    0.774  &   0.862   &  0.816   &   0.890 & \\ \hline
& \textbf{0.805}   &  \textbf{0.195}  &    \textbf{0.809}   &   \textbf{0.805}  &   \textbf{0.805}  &    \textbf{0.890} &  Weighted Avg. \\ \hline
\end{tabular}
\end{table*}

\begin{table*}
\centering
\caption{Detailed Accuracy by Class for Validation Logistic Regression Model}
\label{table:detailedAccuracyValidation}
\begin{tabular}{|c|c|c|c|c|c|c|c|} \hline 
\textbf{Class} & \textbf{TP Rate} & \textbf{FP Rate} & \textbf{Precision} & \textbf{Recall} & \textbf{F-measure} & \textbf{ROC Area} &  \\ \hline
\textbf\textit{scholar} & 0.747  &   0.135  &    0.847  &   0.747  &   0.794   &   0.891 & \\ \hline
\textbf\textit{non-scholar} & 0.865  &   0.253  &    0.773  &   0.865   &  0.817   &   0.891 & \\ \hline
& \textbf{0.806}   &  \textbf{0.194}  &    \textbf{0.810}   &   \textbf{0.806}  &   \textbf{0.805}  &    \textbf{0.891} &  Weighted Avg. \\ \hline
\end{tabular}
\end{table*}

\begin{figure*}
\centering
\includegraphics[height=4in, width=\textwidth]{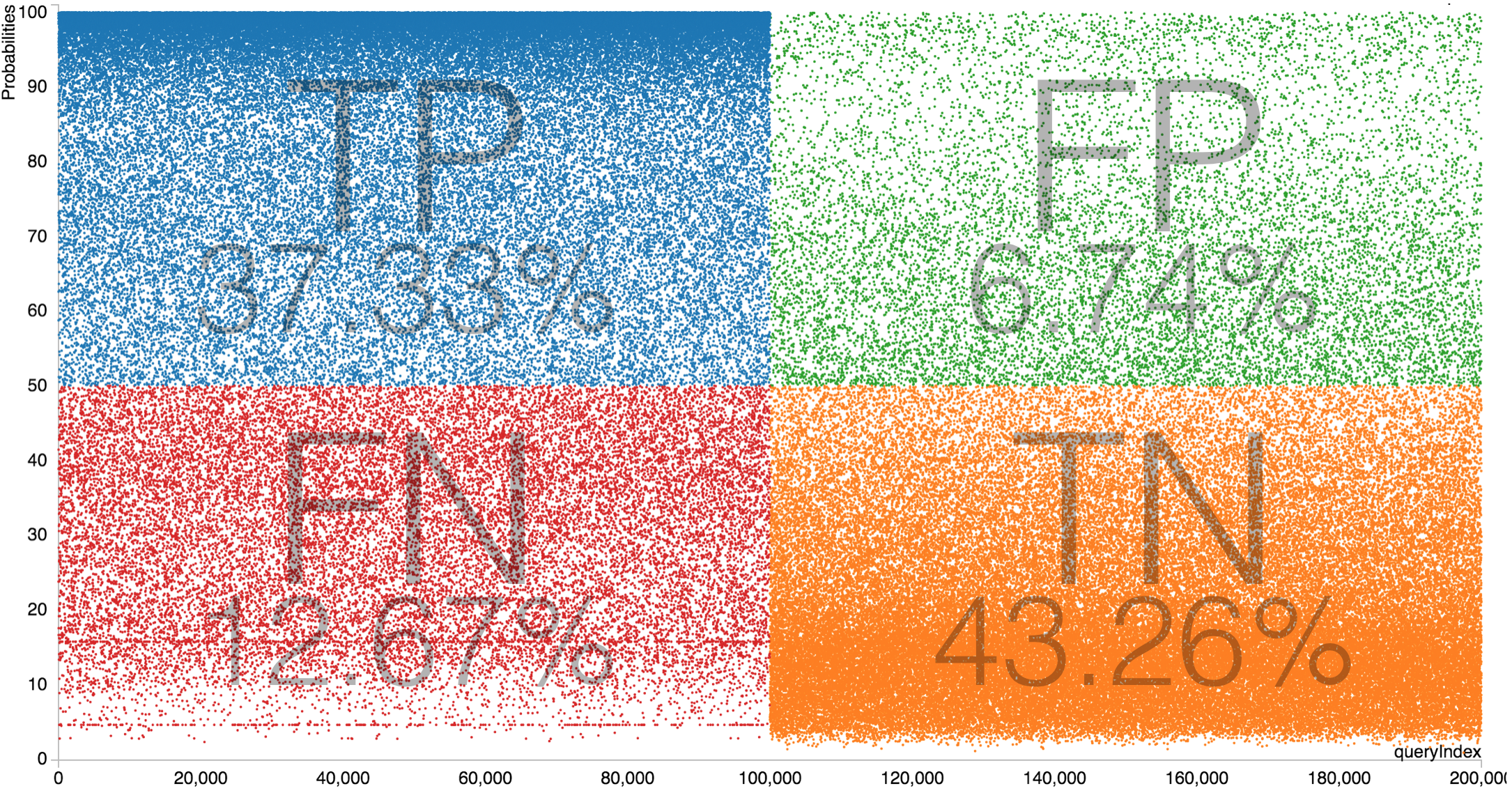}
\caption{Distribution of Validation Model Class Probabilities for 200,000 Queries evenly split across \textit{\textbf{scholar}} and \textit{\textbf{non-scholar}} classes: Blue (Row 1, Col. 1, TP) \textit{\textbf{scholar}}, Orange (Row 2, Col. 2, TN) \textit{\textbf{non-scholar}}, Green (Row 1, Col. 2, FP), Red (Row 2, Col. 1, FN).}
\end{figure*}

\begin{figure*}
\centering
\includegraphics[height=0.8\textheight, width=\textwidth]{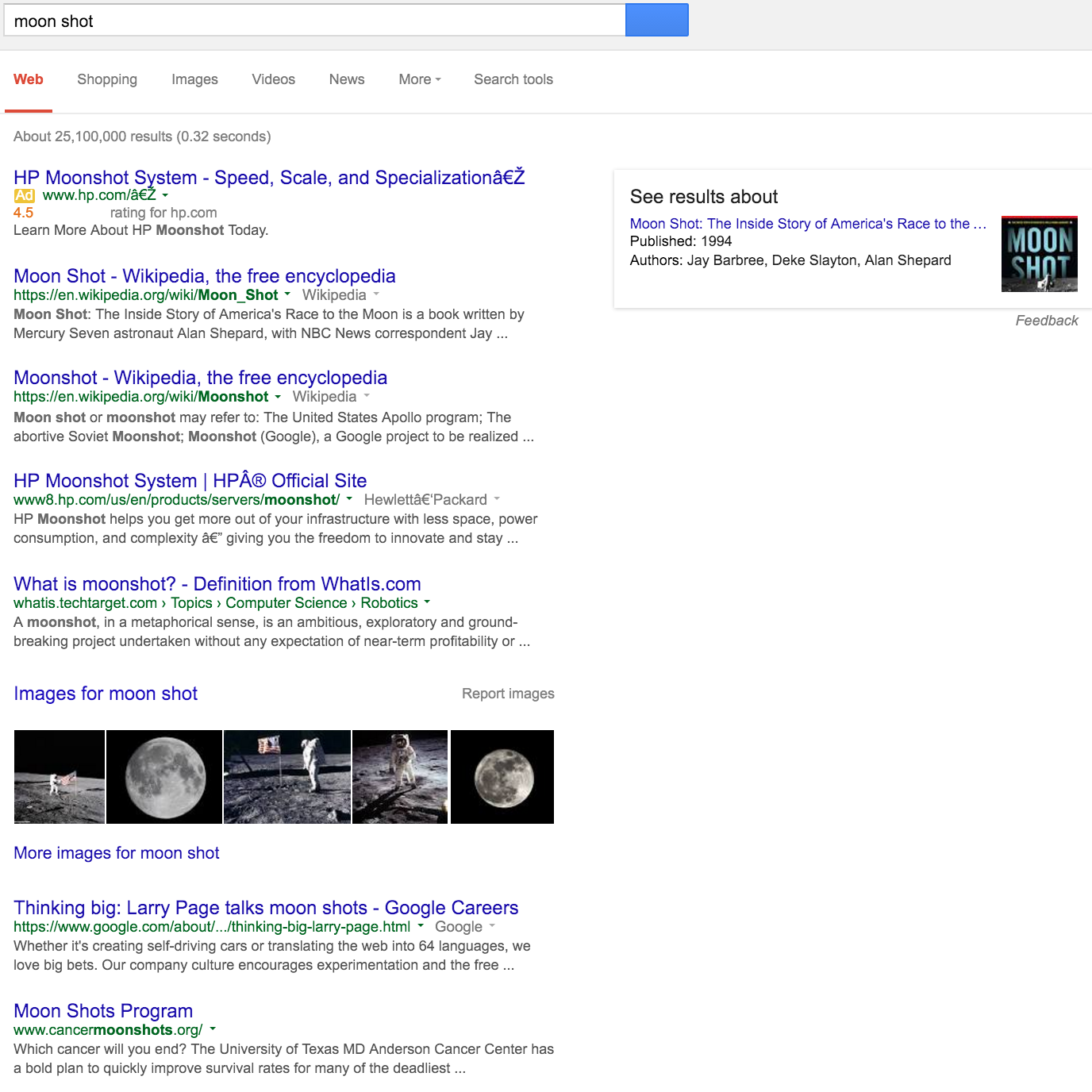}
\caption{Portion of SERP for ``moon shot'' query, used for the example in Section 5.3}
\end{figure*} 

\begin{figure*}
	\centering
	\includegraphics[height=0.75\textheight, width=\textwidth]{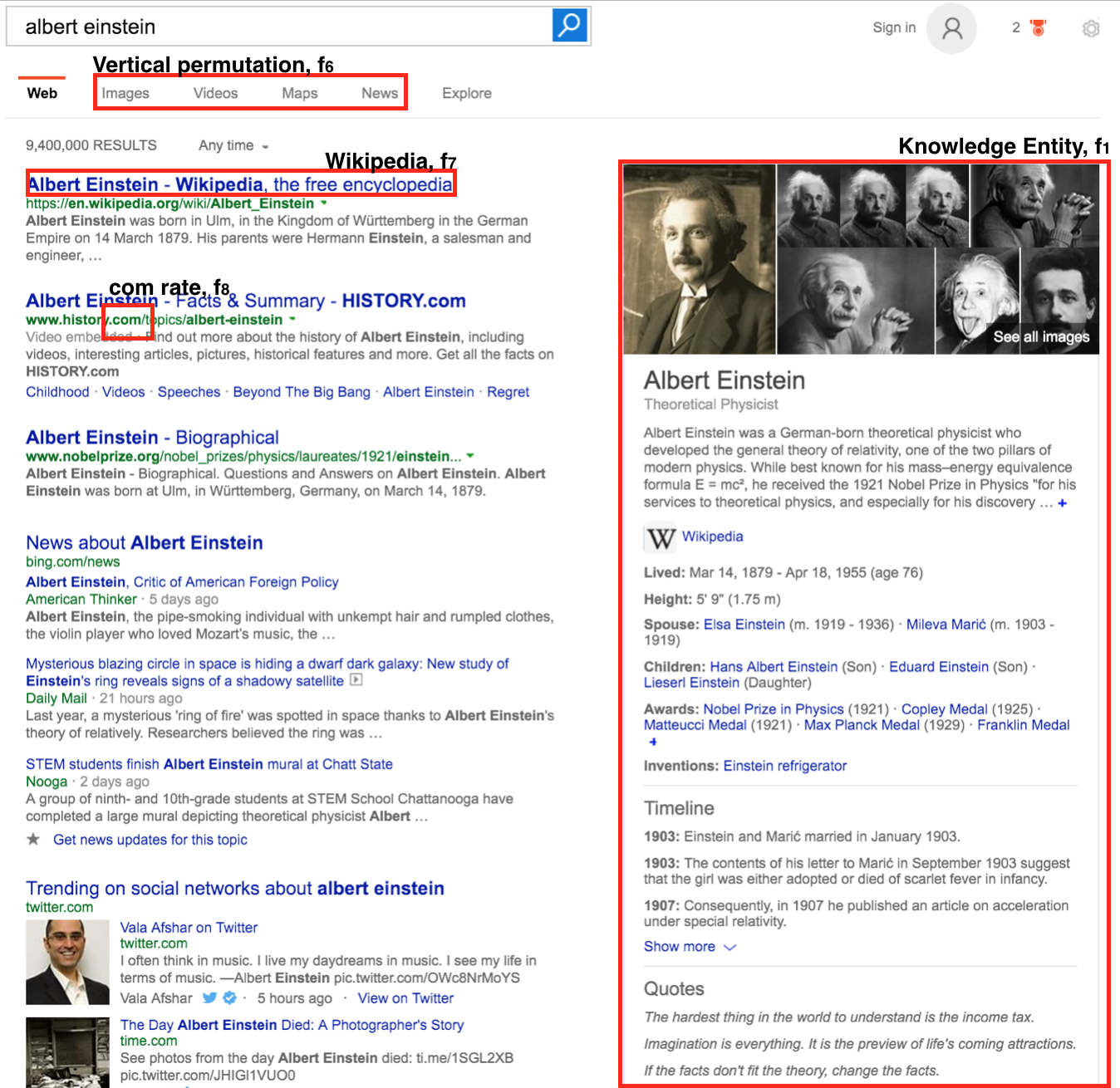}
	\caption{Bing SERP with for ``albert einstein'' query with features $f_1$, $f_6$, $f_7$ and $f_8$. Snapshot date: April 16, 2016.}
\end{figure*}

\begin{enumerate}
	\item \textbf{Issue the query to Google and download SERP (Fig. 7)}
	\item \textbf{Initialize all feature values $f_1$ ... $f_{10}$:}
	\begin{enumerate}
		\item Knowledge Entity, $f_1 = 1$ (Absent)
		\item Images, $f_2 = 0$ (Present)
		\item Google Scholar, $f_3 = 1$ (Absent)
		\item Ad ratio, $f_4 = 0.0833$ (1 ad out of 12 results)
		\item \textit{non-HTML} rate, $f_5 = 0$ (SERP types has no member in set M (Section 4))
		\item Vertical permutation, $f_6 = 275$ (\textit{Shopping - 6, Images - 3, Videos - 7})
		\item Wikipedia, $f_7 = 0$ (Present)
		\item \textit{com} rate, $f_8 = 0.6363$ (7 \textit{com} Tlds out of 11 links)
		\item Maximum Title Dissimilarity, $f_9 = 0.8478$ (See Table ~\ref{table:f9f10}.):

		Given query $q$ = ``moon shot'',

		Fourth SERP title, $t_4$ = ``What is moonshot? - Definition from WhatIs.com'', 

		Maximum SERP title length, $|T|$ = $|t_4|$ = $46$,

		Levinshtein Distance between $q$ and $t_4$, $LD(q, t_4)$ = $39$

		\large
		$f_9$ = $\max_{t_i}\frac{LD(q, t_4)}{|T|}$ \normalsize = $\frac{39}{46} = 0.8478$

		\item Maximum Title Overlap, $f_{10} = 2$ (See Table ~\ref{table:f9f10}.):

		Given query $q$ = ``moon shot'', transformed to query set,
		$q'$ = \{``moon'', ``shot''\}, 

		SERP title 
		$t_1$ = ``Moon Shot - Wikipedia, the free encyclopedia'', 
		transformed to SERP title set, 
		
		$t'_1$ = \{``shot'', ``wikipedia,'', ``-'', ``free'', ``moon'', ``encyclopedia'', ``the''\},

		\large
		$f_{10} = \max_{t'_i} | q' \cap t'_i |$\normalsize

		= 
		\{``moon'', ``shot''\} $\cap$ \{``shot'', ``wikipedia,'', ``-'', ``free'', ``moon'', ``encyclopedia'', ``the''\}
		= $2$

	\end{enumerate}
	\item \textbf{Use Eqn 3. to estimate $g_q$:}

	$
	g_q = 2.7585 + 
	( 0.8266.f_1 ) + 
	( −-1.1664.f_2 ) + 
	( −-2.7413.f_3 ) + 
	( −-1.7444.f_4 ) + 
	( 6.2504.f_5 ) + 
	( −-0.0017.f_6 ) + 
	( −-1.0145.f_7 ) + 
	( −-1.5367.f_8 ) + 
	( 1.8977.f_9 ) + 
	( -0.1737.f_{10} )
	$

	=
	$
	2.7585 + 
	( 0.8266 \times 1 ) + 
	( −-1.1664 \times 0 ) + 
	( −-2.7413 \times 1 ) + 
	( −-1.7444 \times 0.0833 ) + 
	( 6.2504 \times 0 ) + 
	( −-0.0017 \times 275 ) + 
	( −-1.0145 \times 0 ) + 
	( −-1.5367 \times 0.6363 ) + 
	( 1.8977 \times 0.8478 ) + 
	( -0.1737 \times 2 )
	$

	= 0.5147

	\item \textbf{Use the Logistic Regression Hypothesis (Eqn 4.) to estimate the class probability $p_q$:}

	\large$p_q = \frac{e^{g_q}}{1 + e^{g_q}} = \frac{e^{0.5147}}{1 + e^{0.5147}} = 0.6259 $
	\normalsize

	\item \textbf{Classify query:}

	Since $p_q\geq0.5$, domain = \textit{\textbf{scholar}}

\end{enumerate}

\section{Results}

The logistic regression model was built using a 600,000 dataset split evenly across the \textit{\textbf{scholar}} and \textit{\textbf{non-scholar}} classes. The stratified 10-fold cross validation method was used to evaluate the model. The model produced a precision (weighted average over classes) of 0.809 and F-measure of 0.805 (Table ~\ref{table:detailedAccurayLogistic}). The model produced the confusion matrix outlined in Table ~\ref{table:confusionMatrixLogistic}.
\begin{table}
\centering
\caption{Logistic Regression Model Confusion Matrix}
\label{table:confusionMatrixLogistic}
\begin{tabular}{ r|c|c| }
\multicolumn{1}{r}{}
 &  \multicolumn{1}{c}{\textbf{scholar}}
 & \multicolumn{1}{c}{\textbf{non-scholar}} \\
\cline{2-3}
\textbf{scholar} & TP: 224,360 & FP: 41,266 \\
\cline{2-3}
\textbf{non-scholar} & FN: 75,640 & TN: 258,734 \\
\cline{2-3}
\end{tabular}
\end{table}

In order to verify if the model was in fact as good as the preliminary results showed, we validated the model with a fresh ``unseen'' (not used to train/test the classifier) dataset of 200,000 evenly split across both the \textit{\textbf{scholar}} and \textit{\textbf{non-scholar}} class. The validation model yielded a precision of 0.810 (weighted average over classes) and F-measure of 0.805 (Table ~\ref{table:detailedAccuracyValidation}, Fig. 6). The model produced the confusion matrix outlined in Table ~\ref{table:validationModel}.

\begin{table}
\centering
\caption{Validation: Logistic Regression Model Confusion Matrix}
\label{table:validationModel}
\begin{tabular}{ r|c|c| }
\multicolumn{1}{r}{}
 &  \multicolumn{1}{c}{\textbf{scholar}}
 & \multicolumn{1}{c}{\textbf{non-scholar}} \\
\cline{2-3}
\textbf{scholar} & TP: 74,651 & FP: 13,473 \\
\cline{2-3}
\textbf{non-scholar} & FN: 25,349 & TN: 86,527 \\
\cline{2-3}
\end{tabular}
\end{table}

\begin{table*}
\centering
\caption{Some Impure Queries in NTRS (\textit{\textbf{scholar}}) Dataset and AOL (\textit{\textbf{non-scholar}}) Dataset which Contribute to Classification Errors}
\label{table:impureQueries}
\begin{tabular}{|c|c|c|c|c|} 
   \multicolumn{3}{c}{\textbf{Scholar}} & \multicolumn{2}{c}{\textbf{non-Scholar}} \\ \hline
\textbf{Index} & \textbf{Query} & \textbf{Class Probability} & \textbf{Query} & \textbf{Class Probability}  \\ \hline
1 & team and manage  &   0.1325  & team and manage  &   0.1325  \\ \hline
2 & source supplies  &   0.1351  & source supplies  &   0.1351  \\ \hline
3 & solour  &   0.1661  & solour  &   0.1661  \\ \hline
4 & Pinkerton  &   0.1914  & Pinkerton  &   0.1914  \\ \hline
5 & prying  &   0.2075  & prying  &   0.2075  \\ \hline
6 & screen saver  &   0.2726  & screen saver  &   0.2726  \\ \hline
7 & race car tests  &   0.2830  & race car tests  &   0.2830  \\ \hline
8 & guaranteed funding  &   0.2888  & guaranteed funding  &   0.2888  \\ \hline
9 & fix  &   0.4927  & fix  &   0.4927  \\ \hline
\end{tabular}
\end{table*}

\section{Discussion}

Google is essentially a black box. Callan \cite{callan2002distributed} explored how Search Engines search a number of databases, and merge results returned by different databases by ranking them based on their ability to satisfy the query. Similarly, Murdock and Lalmas \cite{murdock2008workshop} explored aggregated search, addressing how to select, rank and present resources in a manner that information can be found efficiently and effectively. Arguello et al. \cite{arguello2009sources} addressed the problem of vertical selection which involves predicting relevant verticals for user queries issued to the Search Engine's main web search page. Even though these methods (\cite{callan2002distributed}, \cite{murdock2008workshop}, and \cite{arguello2009sources}) may shed some light about some operations associated with Search Engines, we still do not know about all the mechanism of the Google search, but we can see what it returns. Also, instead of replicating the methods discussed in Section 2 and methods such as \cite{callan2002distributed}, \cite{murdock2008workshop}, and \cite{arguello2009sources}, we leverage the results produced by Search Engines.

Before building our dataset, we had to be able to detect and extract the features from SERPs. This was achieved by scraping the Google SERP. While identifying the features was an easy task, the limitation of scraping is that if the HTML elements (e.g., tags, attributes) of the SERP change, the implementation might need to change too. However, based on empirical evidence, SERPs tend to change gradually. But if they do, for example, if Google adds a new Vertical or a new source of evidence, this could result in a new feature for our method, therefore our implementation \cite{QueryClassification}, has to be updated. Our method processes Google SERPs, but it could easily be adapted to other SERPs such as Bing. Fig. 8 outlines some features from the Google SERP that carry over to Bing.

Before building our model, we visualized the datasets at our disposal and performed information analysis (information gain/gain ratio - Table ~\ref{tab:completeListFeatures}) in order identify features which could contribute discrimination information for our classifier. After identifying 10 features (Table ~\ref{tab:completeListFeatures}), we trained and evaluated a logistic regression model on the dataset and achieved a precision of 0.809 and F-measure of 0.805. Our method is not without limitations. For example, the datasets are presumed ``pure.'' But this is not the case since Scholar queries exist in the non-Scholar dataset and vice versa, thus contribute to classification errors. However, it should be noted that our model performed very well even though it was trained on an ``impure'' dataset (e.g., Table ~\ref{table:impureQueries}).

The NTRS dataset (\textit{\textbf{scholar}} class) had queries which could be easily identified as \textit{\textbf{non-scholar}} queries. For example, consider Table ~\ref{table:impureQueries} which presents some queries found in the NTRS dataset with questionable membership, thus contributing to classification errors. The low class probabilities could signal that the classifier had little ``doubt'' when it placed the \textit{\textbf{non-scholar}} label (correct based on human judgement) on such queries, but was penalized due to the presence of these queries in the \textit{\textbf{scholar}} dataset. Therefore, our Classifier could be improved if the ``impure'' queries and their respective SERPs are removed.

Similar to the NTRS dataset, the AOL 2006 dataset (\textit{\textbf{non-scholar}} class) also had queries which could be easily identified as \textit{\textbf{scholar}} queries. For example, Table ~\ref{table:impureQueries} presents some queries found in the AOL dataset with questionable membership, thus contributing to classification errors. It is clear from the high class probabilities, that the classifier had little ``doubt'' when it placed the \textit{\textbf{scholar}} label (correct based on human judgement) on such queries, but was penalized due to the presence of these queries in the \textit{\textbf{non-scholar}} dataset. The python open source code for our implementation and the NTRS dataset is available for download at \cite{QueryClassification}.

\section{Conclusions}
In order to route a query to its appropriate digital library, we established query understanding, (thus domain classification) as a means toward deciding if a query is better served by a given digital library. The problem of domain classification is not new, however the state of the art consists of a spectrum of solutions which process the query unlike our method which leverages quality SERPs. We define a set of ten features in SERPs that indicate if the domain of a query is scholarly or not. Our logistic regression classifier was trained and evaluated (10-fold cross validation) with a SERP corpus of 600,000 evenly split across the \textit{\textbf{scholar}}, \textit{\textbf{non-scholar}} classes, yielding a precision of 0.809 and F-measure of 0.805. The performance of the classifier was further validated with a SERP corpus of 200,000 evenly split across both classes yielding approximately the same precision and F-measure as before. Even though our solution to the domain classification problem targets two domains (\textit{\textbf{scholar}} and \textit{\textbf{non-scholar}}), our method can be easily extended to accommodate more domains once discriminative and informative features are identified.


%
\bibliographystyle{abbrv}
\bibliography{sigproc}  

\begin{thebibliography}{10}

\bibitem{arguello2009sources}
J.~Arguello, F.~Diaz, J.~Callan, and J.-F. Crespo.
\newblock Sources of evidence for vertical selection.
\newblock In {\em Proceedings of the 32nd international ACM SIGIR Conference on
  Research and Development in Information Retrieval}, pages 315--322. ACM,
  2009.

\bibitem{callan2002distributed}
J.~Callan.
\newblock Distributed information retrieval.
\newblock In {\em Advances in Information Retrieval}, pages 127--150. Springer,
  2002.

\bibitem{Gravano:reasoning}
L.~Gravano, V.~Hatzivassiloglou, and R.~Lichtenstein.
\newblock Categorizing web queries according to geographical locality.
\newblock In {\em Proceedings of the Twelfth International Conference on
  Information and Knowledge Management}, pages 325--333. ACM, 2003.

\bibitem{hall2009weka}
M.~Hall, E.~Frank, G.~Holmes, B.~Pfahringer, P.~Reutemann, and I.~H. Witten.
\newblock The {WEKA} data mining software: an update.
\newblock {\em ACM SIGKDD Explorations Newsletter}, 11(1):10--18, 2009.

\bibitem{hu2009understanding}
J.~Hu, G.~Wang, F.~Lochovsky, J.-t. Sun, and Z.~Chen.
\newblock Understanding user's query intent with wikipedia.
\newblock In {\em Proceedings of the 18th International Conference on World
  Wide Web}, pages 471--480. ACM, 2009.

\bibitem{jansen2008determining}
B.~J. Jansen, D.~L. Booth, and A.~Spink.
\newblock Determining the informational, navigational, and transactional intent
  of web queries.
\newblock {\em Information Processing \& Management}, 44(3):1251--1266, 2008.

\bibitem{larose2006data}
D.~T. Larose.
\newblock {\em Data mining methods \& models}.
\newblock John Wiley \& Sons, 2006.

\bibitem{lee2005automatic}
U.~Lee, Z.~Liu, and J.~Cho.
\newblock Automatic identification of user goals in web search.
\newblock In {\em Proceedings of the 14th International Conference on World
  Wide Web}, pages 391--400. ACM, 2005.

\bibitem{liu2013query}
J.~Liu, P.~Pasupat, Y.~Wang, S.~Cyphers, and J.~Glass.
\newblock Query understanding enhanced by hierarchical parsing structures.
\newblock In {\em Automatic Speech Recognition and Understanding (ASRU), 2013
  IEEE Workshop on}, pages 72--77. IEEE, 2013.

\bibitem{mccullagh1989generalized}
P.~McCullagh and J.~A. Nelder.
\newblock {\em Generalized linear models}, volume~37.
\newblock CRC press, 1989.

\bibitem{murdock2008workshop}
V.~Murdock and M.~Lalmas.
\newblock Workshop on aggregated search.
\newblock In {\em ACM SIGIR Forum}, volume~42, pages 80--83. ACM, 2008.

\bibitem{ntrs:ir}
M.~L. Nelson, G.~L. Gottlich, D.~J. Bianco, S.~S. Paulson, R.~L. Binkley, Y.~D.
  Kellogg, C.~J. Beaumont, R.~B. Schmunk, M.~J. Kurtz, A.~Accomazzi, and
  O.~Syed.
\newblock The {NASA} technical report server.
\newblock {\em Internet Research}, 5(2):25--36, 1995.

\bibitem{QueryClassification}
A.~Nwala.
\newblock {Query Classification Source Code}.
\newblock \url{https://github.com/oduwsdl/QueryClassification}, 2016.

\bibitem{Pass:2006:PS:1146847.1146848}
G.~Pass, A.~Chowdhury, and C.~Torgeson.
\newblock A picture of search.
\newblock In {\em Proceedings of the 1st International Conference on Scalable
  Information Systems}, 2006.

\bibitem{shen2005q}
D.~Shen, R.~Pan, J.-T. Sun, J.~J. Pan, K.~Wu, J.~Yin, and Q.~Yang.
\newblock {Q2C@UST}: our winning solution to query classification in {KDDCUP}
  2005.
\newblock {\em ACM SIGKDD Explorations Newsletter}, 7(2):100--110, 2005.

\bibitem{sugiura2000query}
A.~Sugiura and O.~Etzioni.
\newblock Query routing for web search engines: Architecture and experiments.
\newblock {\em Computer Networks}, 33(1):417--429, 2000.

\bibitem{wang2009semi}
Y.-Y. Wang, R.~Hoffmann, X.~Li, and J.~Szymanski.
\newblock Semi-supervised learning of semantic classes for query understanding:
  from the web and for the web.
\newblock In {\em Proceedings of the 18th ACM Conference on Information and
  Knowledge Management}, pages 37--46. ACM, 2009.

\bibitem{survey}
C.~Wolff, A.~B. Rod, and R.~C. Schonfeld.
\newblock {Ithaka S+R US Faculty Survey 2015}.
\newblock
  \url{http://www.sr.ithaka.org/wp-content/uploads/2016/03/SR_Report_US_Faculty_Survey_2015040416.pdf},
  2016.

\bibitem{Jingbo:label}
J.~Yu and N.~Ye.
\newblock Automatic web query classification using large unlabeled web pages.
\newblock In {\em Web-Age Information Management, 2008. WAIM'08. The Ninth
  International Conference on}, pages 211--215. IEEE, 2008.

\end{thebibliography}
%
%




\end{document}